# Topological Defects Induced High-Spin Quartet State in Truxene-Based Molecular Graphenoids

Can Li,[†][a] Yu Liu,[†][a] Yufeng Liu,[†][a] Fu-Hua Xue,[2][b] Dandan Guan,[a,c,d] Yaoyi Li,[a,c,d] Hao Zheng,[a,c,d] Canhua Liu,[a,c,d] Jinfeng Jia,[a,c,d] Pei-Nian Liu,[b] Deng-Yuan Li,*[b] Shiyong Wang*[a,c,d]

[a]  C. Li, Y. Liu, Y. Liu, Dr. D. Guan, Dr. Y. Li, Prof. H. Zheng, Prof. C. Liu, Prof. J. Jia, Prof. S. Wang
     Key Laboratory of Artificial Structures and Quantum Control (Ministry of Education), Shenyang National Laboratory for Materials Science, School of Physics and Astronomy, Shanghai Jiao Tong University, Shanghai 200240, China; E-mail: shiyong.wang@sjtu.edu.cn

[b]  F.-H. Xue, Prof. P.-N. Liu, Dr. D.-Y. Li
     Key Laboratory for Advanced Materials and Feringa Nobel Prize Scientist Joint Research Center, Frontiers Science Center for Materiobiology and Dynamic Chemistry, School of Chemistry and Molecular Engineering, East China University of Science & Technology, 130 Meilong Road, Shanghai, 200237, China; E-mail: dengyuanli@ecust.edu.cn

[c]  Dr. D. Guan, Dr. Y. Li, Prof. H. Zheng, Prof. C. Liu, Prof. J. Jia, Prof. S. Wang
     Tsung-Dao Lee Institute, Shanghai Jiao Tong University, Shanghai, 200240, China

[d]  Dr. D. Guan, Dr. Y. Li, Prof. H. Zheng, Prof. C. Liu, Prof. J. Jia, Prof. S. Wang
     Shanghai Research Center for Quantum Sciences, Shanghai 201315, China

[†]  These authors contributed equally to this work.

**Abstract**

Topological defects in graphene materials introduce exotic properties which are absent in their defect-free counterparts with both fundamental importance and technological implications. Although individual topological defects have been widely studied, collective magnetic behaviors originating from well-organized multiple topological defects remain a great challenge. Here, we studied the collective magnetic properties originating from three pentagon topological defects in truxene-based molecular graphenoids by using scanning tunneling microscopy and non-contact atomic force microscopy. Unpaired π electrons are introduced into the aromatic topology of truxene molecular graphenoids one by one by dissociating hydrogen atoms at the pentagon defects via atom manipulation. Scanning tunneling spectroscopy measurements together with density functional theory calculations suggest that the unpaired electrons are ferromagnetically coupled, forming a collective high-spin quartet state of $S=3/2$. Our work demonstrates that the collective spin ordering can be realized through engineering regular patterned topological defects in molecular graphenoids, providing a new platform for designer one-dimensional ferromagnetic spin chains and two-dimensional ferromagnetic networks.

**Introduction**

Topological defects widely exist in graphene materials and locally change their intrinsic physical and chemical properties. Examples include $sp^3$ carbons, vacancies, carbon tetragons, pentagons, heptagons, dislocations, and grain boundaries, which introduce intriguing electronic, magnetic, optical, and mechanical properties with implications for sensors, electronics, spintronics, and quantum technologies.[1-13] Topological defects usually break the sublattice symmetry of graphene and introduce unconventional π-electron magnetism, which has properties that differ from *d/f* electron magnetism.[14] In the past decade, intense research efforts have been devoted to studying such unconventional magnetism in graphene materials. Although many graphene systems exhibit signatures of the presence of $sp^2$ carbon magnetism, it remains elusive to reveal collective magnetic behaviors originating from regular patterned topological defects. This is mainly because the topological defects are usually randomly interspersed in graphene lattice, and how to arrange the topological defects in a controllable way remains challenging by using traditional top-down methods. Additionally, as the magnetism of graphene-based materials depends crucially on their atomic structure and surrounding environments, it not only requires the ability to fabricate samples with atomic precision but needs techniques to characterize them down to the single-chemical-bond level.

Recently, atomically precise segments of graphene, also named nanographenes or molecular graphenoids, with carefully designed topological defects have been reported by taking advantage of recent advances in on-surface synthesis and scanning probe techniques.[15-25] On-surface synthesis approach composes of precursor synthesis in solution and subsequent surface-assisted reactions of precursor building blocks, which permits the synthesized graphene nanostructures with atomic precision and wide tunability.[26-32] After synthesis, the structural and electronic properties of molecular graphenoids can be directly characterized by surface techniques, mainly by a combined non-contact atomic force microscopy (nc-AFM) and scanning tunneling microscopy (STM) system with the single-chemical-bond spatial resolution.[33] Recent studies using such bottom-up surface techniques have revealed that graphene nanostructures with topological defects host interesting localized electronic states, tunable band gap, and magnetic properties.[21,22,34,35] Compared to the planar configuration of polyaromatic hydrocarbons with purely hexagonal rings, molecular graphenoids with topological defects usually feature a structural curvature.[15,17,18] Such structural deformation is usually accompanied by a strong electron coupling with the underneath metal substrate, hindering the detection of their intrinsic magnetic properties. Thus, characterization of collective magnetic behaviors originating from multiple topological defects in molecular graphenoids remains a great challenge.

Here, we report the study of collective high-spin states of a truxene molecule (10,15-dihydro-5H-diindeno[1,2-a:1',2'-c]fluorene), which is a heptacyclic polyarene structure containing three pentagons with one $sp^3$ carbon vertice in per pentagon ring.[36] Using atom manipulation, three hydrogen atoms at the $sp^3$ carbons are controllably dissociated one by one, transforming the $sp^3$ carbons into $sp^2$ carbons. Due to their fully aromatic character, up to three unpaired π electrons are introduced into the aromatic topology, forming a high-spin quartet state as suggested by density functional theory (DFT) calculations. High-resolution nc-AFM imaging reveals that the dehydrogenated truxene molecules (DTrs) adopted a curved structure on Au(111) with the pentagons having lower adsorption heights. Due to electronic coupling with underneath Au(111), the resulting high-spin states of DTrs are quenched on Au(111) as revealed by scanning tunneling spectroscopy (STS) measurements. To probe their intrinsic electronic structure, we deposit truxene molecules in-situ on an insulating NaCl/Au(111) substrate, and dissociate hydrogens one by one forming DTrs. By sharp contrast with Au(111), singly occupied/unoccupied molecular orbitals (SOMOs/SUMOs) are resolved for DTrs on NaCl film, confirming the presence of high-spin states. Our results may be further extended to realize one-dimensional ferromagnetic chains and two-dimensional ferromagnetic frameworks.

**Results and Discussion**

According to Lieb's theorem and Ovchinnikov's rule, molecular graphenoids with sublattice imbalance in the bipartite honeycomb lattice host tunable magnetic ground states.[14] The spin quantum number $S$ of the ground state is given by $S=(N_A-N_B)/2$, where $N_A(N_B)$ is the number of carbon atoms in the $A(B)$ sublattice of the graphene honeycomb lattice. As shown in Figure 1a, the application of Lieb's theorem predicates triangular zigzag nanographenes, such as triangulene and its larger homologues, host tunable magnetic ground states. Experimentally, [2]-triangulene (3 fused rings, $S=1/2$) and [3]-triangulene (6 fused rings, $S=1$) have been synthesized in solution and their open-shell character has been confirmed by electron spin resonance spectroscopy;[37-40] [4]-triangulene (10 fused rings, $S=3/2$), [5]-triangulene (15 fused rings, $S=2$), and [7]-triangulene (28 fused rings, $S=3$) have been obtained by on-surface synthesis.[41-44] In the triangulene system, the size of triangulene molecules increases quickly with the spin quantum number $S$, where the total required number of fused benzene rings in triangulene is given by $N=(S+1)\times(2S+1)$. Thus, it is challenging to fabricate triangulenes with spin quantum number *S* larger than 3 due to difficulties encountered in the synthesis and sublimation of large molecular precursors onto surfaces.

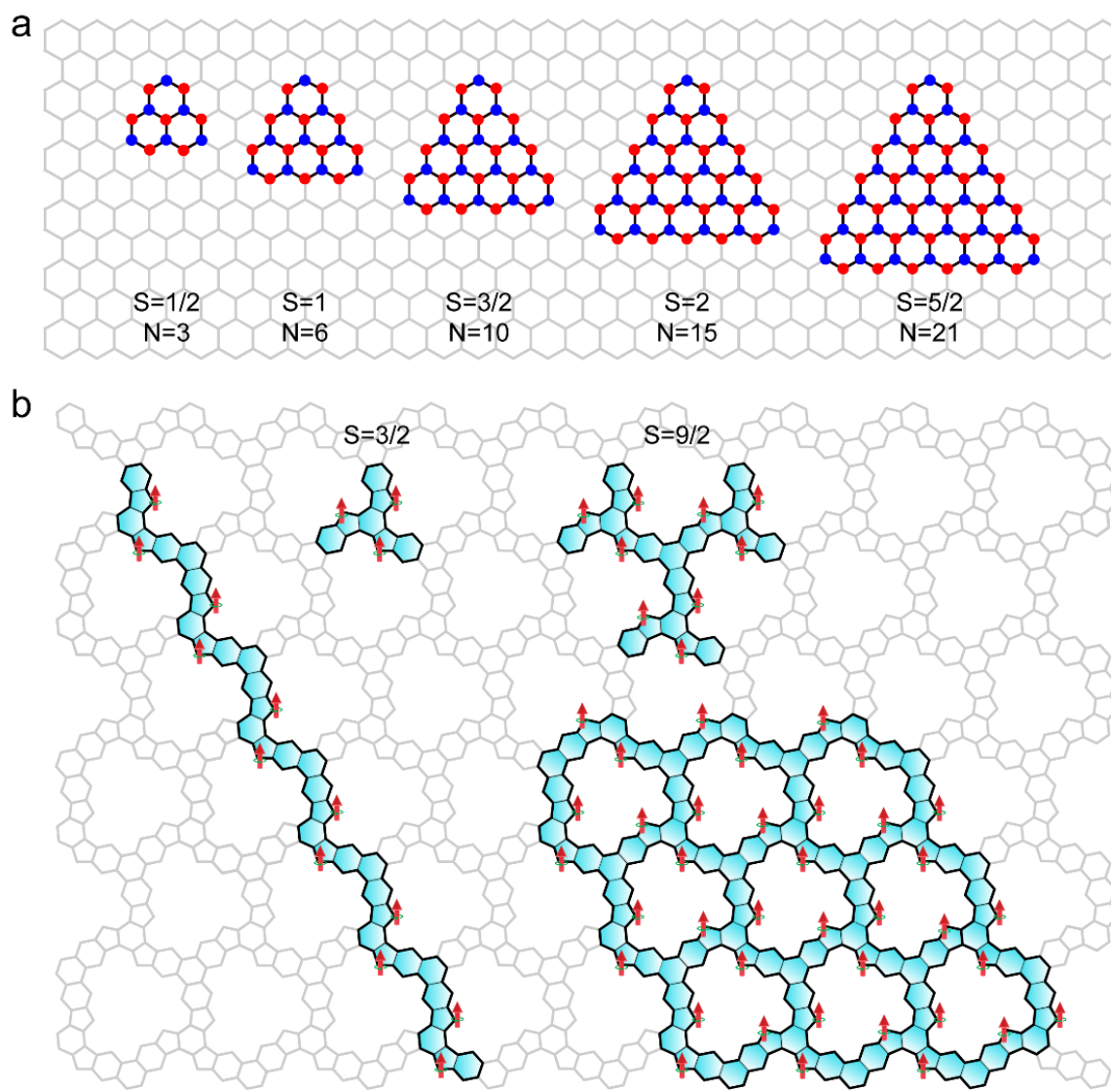

**Figure 1.** High-spin states in nanographene systems. (a) Magnetic molecules of triangluene-based system with spin *S*, while the *N* indicates the number of fused benzene rings in triangluene-based molecules. (b) Conceptual routes towards high-spin state molecules, one-dimensional ferromagnetic spin chain, and two-dimensional ferromagnetic network based on truxene.

In contrary to sublattice imbalance, topological defect is much more efficient to introduce unpaired electrons in molecular graphenoids. The incorporation of one pentagon ring in molecular graphenoids can introduce one unpaired π electron. It is, in principle, possible to efficiently build high spin molecules by periodically positioning electronically coupled topological defects. We searched for the molecular database and found the dehydrogenated truxene molecule (containing three pentagons connected at the *1,3,5*-side of the central benzene ring) is an ideal platform to achieve high spin states due to its C3 symmetry as confirmed by DFT calculations (Figure 2). More interestingly, extended graphenoids using dehydrogenated truxene building blocks host highly tunable spin states. For example, a molecular graphenoid with three truxene molecular building blocks linked through the *1,3,5*-side of a central benzene ring has a ground state spin quantum number of *S*=9/2 (Figure 1b and 2c), with only 22 fused rings much less than those required in [10]-triangulene (55 fused rings). Except for designer high spin molecules, the extended one-dimensional dehydrogenated truxene polymer and two-dimensional network would host intriguing one-dimensional ferromagnetism and two-dimensional ferromagnetism, with wide tunability for exploring low-dimensional quantum magnetism (Figure 1b). As a first step toward these lines, we limit our discussion to the study of the magnetic ground states of individual truxene building blocks in this work.

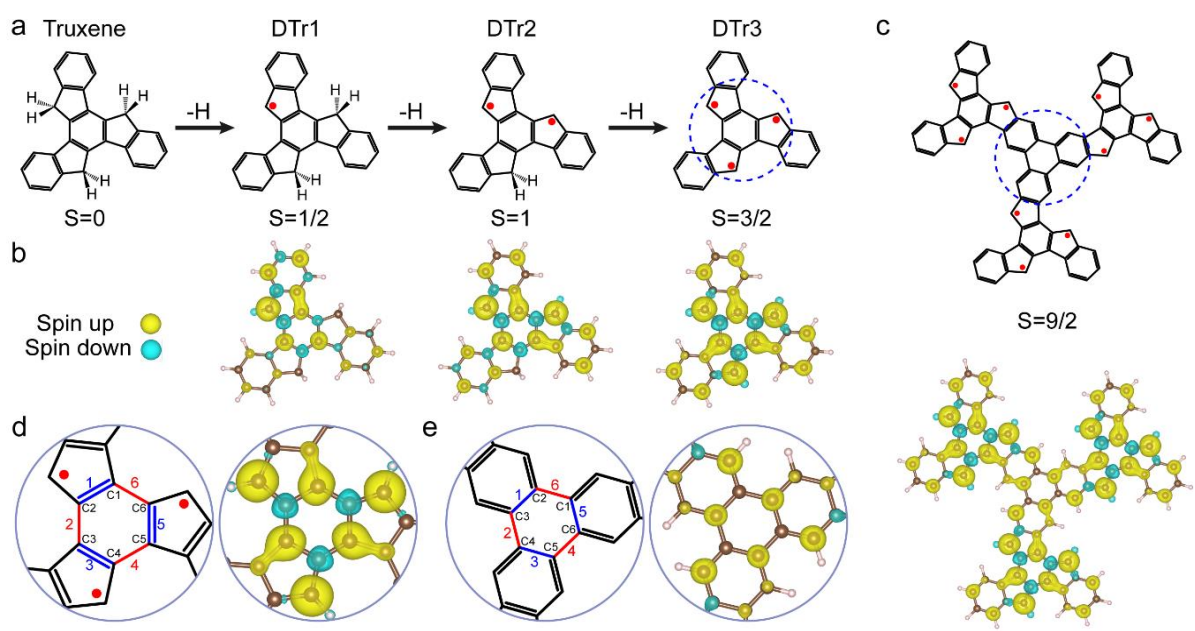

**Figure 2.** Realization of high-spin states in truxene-based frameworks. (a) Chemical structures of dehydrogenated truxene molecules. (b) Spin-density distributions of dehydrogenated truxene molecules corresponding to (a), respectively. (c) High-spin state of *S*=9/2 in a truxene-based trimer. (d) Zoom-in chemical structure and spin density distributions of DTr3 on the center benzene ring labelled by blue dash circle in (a). (e) Zoom-in chemical structure and spin density distributions of DTr3 trimer on the center benzene ring labelled by blue dash circle in (c).

Figure 2 depicts the scheme to realize high-spin systems using truxene molecules. Through STM tip-induced atom manipulation or thermal annealing, the hydrogens at the *sp³* carbon sites of truxene molecular precursor can be dissociated one by one, introducing one, two, and three unpaired π electrons inside the molecule. We refer to the dehydrogenated truxene molecules as DTr1, DTr2, DTr3 for convenience, indicating the presence of one, two, and three unpaired π electrons, respectively (Figure2a). DFT calculations (Figure 2b) predicate that the unpaired electrons are ferromagnetically coupled together, forming high-spin triplet and quartet states for DTr2 and DTr3, respectively. The physics behind such ferromagnetic coupling is attributed to the minimization of on-site Coulomb repulsion in molecular graphenoids with C3 symmetry.[10] Since the three pentagons connect to the *1,3,5*-side of the central benzene ring, the spin-up density prefers to reside on one sublattice (*1,3,5*-carbon sites) and the spin-down density on the other sublattice (*2,4,6*-carbon sites) of the central benzene ring to minimize Coulomb repulsion (the detailed spin density distribution in Figure 2d). In contrast, if the pentagons connect to the *1,4*-side of the central benzene ring, the unpaired electrons will be antiferromagnetically coupled for minimizing Coulomb repulsion.[25,45,46] Figure 2c presents the DFT-calculated spin density distribution of a truxene trimer. Owing to its C3 symmetry, all the unpaired electrons are ferromagnetically coupled together, forming a high spin state with $S=9/2$.

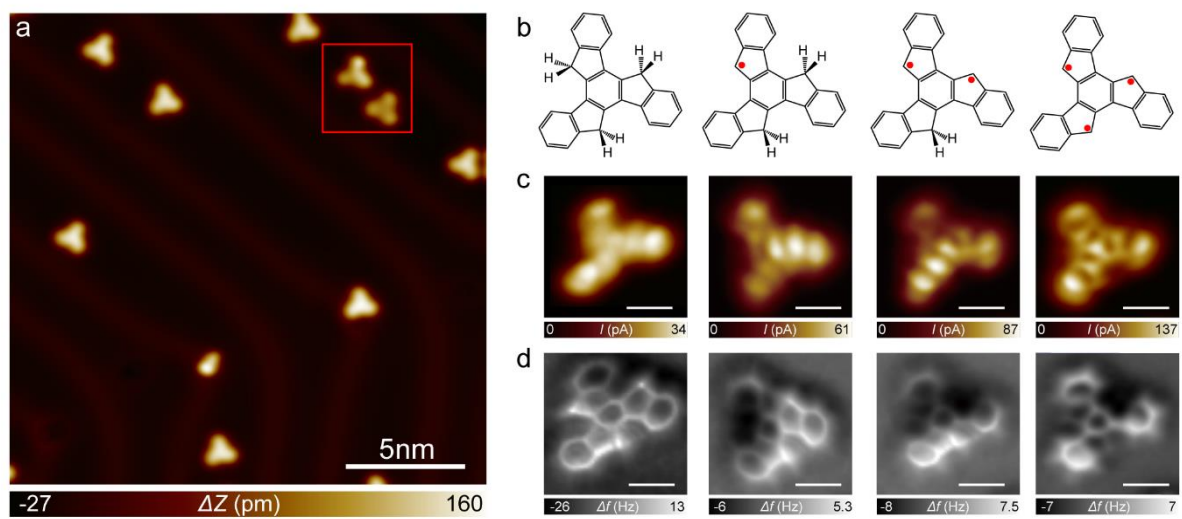

**Figure 3.** Structural properties of dehydrogenated truxene molecular graphnoids on Au(111). (a) Large-scale STM image (Bias voltage: 0.5 V, Current setpoint: 50 pA; Scale bar: 5 nm). Two molecules as marked by red square are dehydrogenated truxenes. (b-d) From top to down: chemical structures, constant-height current images (Bias voltage: 3 mV; Scale bars: 500 pm)

and nc-AFM images (Resonant frequency: 26 kHz, Oscillation amplitude: 80 pm; Scale bars: 500 pm) of truxene precursor, DTr1, DTr2, and DTr3.

The electronic properties of DTrs on a Au(111) substrate were characterized by high-resolution scanning probe microscopy as shown in Figure 3. Truxene precursor was synthesized in solution (Figure S1) and thermally deposited on Au (111) (the details in Supporting Information). The sample was then transferred into a low-temperature STM stage holding at 4 K for characterization. A large-scale STM image in Figure 3a reveals a triangular topology of truxene molecules on Au(111). High-resolution nc-AFM imaging resolves the chemical structure of truxene precursor by functionalizing the tip with a CO molecule, with the $sp^3$ carbons being imaged as a local bright dot since there is one more hydrogen pointing up from $sp^3$ carbons (Figure 3d). With tip-induced atom manipulation, one hydrogen at $sp^3$ carbon was dissociated through positioning the tip above the $sp^3$ carbon and slowly ramping bias voltage to 3.2 V (Figure S2). After dehydrogenation, the DTr1 molecule features a curvature on Au(111), with the dehydrogenated pentagon getting closer to the substrate and the opposite side tilting up (nc-AFM image in Figure 3d). We further dissociated the other two hydrogens one by one and monitored their structural change. Similar observations have been made on all dehydrogenated pentagons, which have a lower adsorption height than surrounding benzene rings. As shown in Figure 3c, constant-height current images taken at 3 mV resolve some fine electronic features originating from the highest occupied molecular orbital (Figure S3). Scanning tunneling differential conductance spectroscopy (dI/dV) measurements have been performed to resolve the detailed electronic properties. For low-energy dI/dV measurements, we cannot obtain any signatures of Kondo resonances for all DTrs on Au(111) (Figure S4), an effect due to the screening of net spins by conduction electrons of Au(111).[47] The absence of Kondo resonance may be due to either the absence of net spins in DTrs or the Kondo temperature is much lower than 4 K. To clarify the physics behind, we performed high energy resolution dI/dV spectra (Figure S5) and recorded the corresponding dI/dV mappings of DTr3. If there indeed exist net spins in DTrs but with much lower kondo temperature on Au(111), the SOMOs resided by the unpaired π electrons should be observed. However, the SOMOs/SUMOs have not been detected for DTr3 on Au(111) (Figure S3). Interestingly, all the observed orbitals can be reproduced by using molecular orbitals of the undehydrogenated truxene precursor, indicating the radicals are quenched on Au(111).

To electronically decouple DTrs away from Au(111) metal substrate, we deposit truxene precursors *in situ* on a NaCl/Au(111) substrate held at 8 K. The NaCl salts were thermally deposited on Au(111) held at room temperature, forming self-assembled NaCl monolayer and bilayer islands. The thin NaCl film has been demonstrated to electronically decouple single molecules away from the underneath metal substrate efficiently.[40,48] Upon in-situ deposition, some truxene molecules directly adsorb on NaCl islands (Figure S6). Similar to the procedures on Au(111), we dissociated hydrogens one by one away from the $sp^3$ carbon sites and monitored their electronic properties. Figure 4 depicts the electronic properties of DTrs on a bilayer NaCl island. DFT-calculated energy spectra indicate that DTr1 hosts a SOMO/SUMO, with a doublet ground state of $S=1/2$. The intrinsic electronic properties of DTr1 on NaCl have been observed experimentally. As shown in Figure 4c, dI/dV spectra taken at different positions of DTr1 resolve the predicted SOMO and SUMO resonances at -1 V and 1.5 V. STM images in Figure 4d taken at these two energies reveal the orbital shapes of SOMO and SUMO, in consistent with STM simulations based on SOMO and SUMO. The presence of SOMO and SUMO confirms the open-shell electronic structure of DTr1 with a net spin of $S=1/2$. DFT-calculated energy spectra of DTr2 in Figure 4b reveal two SOMOs/SUMOs, with a triplet ground state of $S=1$. The obital shapes of two SOMOs/SUMOs are shown in Figure S7, with one orbital having quenched density of state at the undehydrogenated corner. Such singly occupied states have been confirmed by dI/dV spectra in Figure 4c by showing two resonance peaks at -1 V and 1 V, respectively. STM images in Figure 4d taken at these two resonances reveal that the SOMOs and SUMOs have quenched intensity at the undehydrogenated corner, which is reproduced by the simulated STM images in Figure 4e. Additionally, we exclude the possibility that the DTr2 may host a closed-shell electronic structure by comparing its frontier orbitals with experiments (Figure S7). The observation of SOMOs and SUMOs together with their asymmetric spatial distribution confirms the triplet high spin state of DTr2 on NaCl. As shown in Figure 4b, fully dehydrogenated DTr3 is predicted to host three SOMOs/SUMOs, with a high-spin quartet state. Unlike DTr1 and DTr2, all the SOMOs and SUMOs are with the delocalized density of states over the entire molecule (Figure S8). As shown in Figure 4c, dI/dV measurements reveal the presence of SOMOs and SUMOs near -1 V and 1 V. A three-fold symmetric spatial distribution of SOMOs and SUMOs has been confirmed by STM imaging in Figure 4d, in agreement with STM simulations. Additionally, the possibility of DTr3 with a ground state of $S=1/2$ is excluded by comparing its frontier orbitals with experiments (Figure S8). The resolved intrinsic electronic properties of DTr2 and DTr3 on NaCl suggest the presence of collective high-spin triplet and quartet states in truxene-based molecular graphenoids with two and three pentagon defects, respectively.

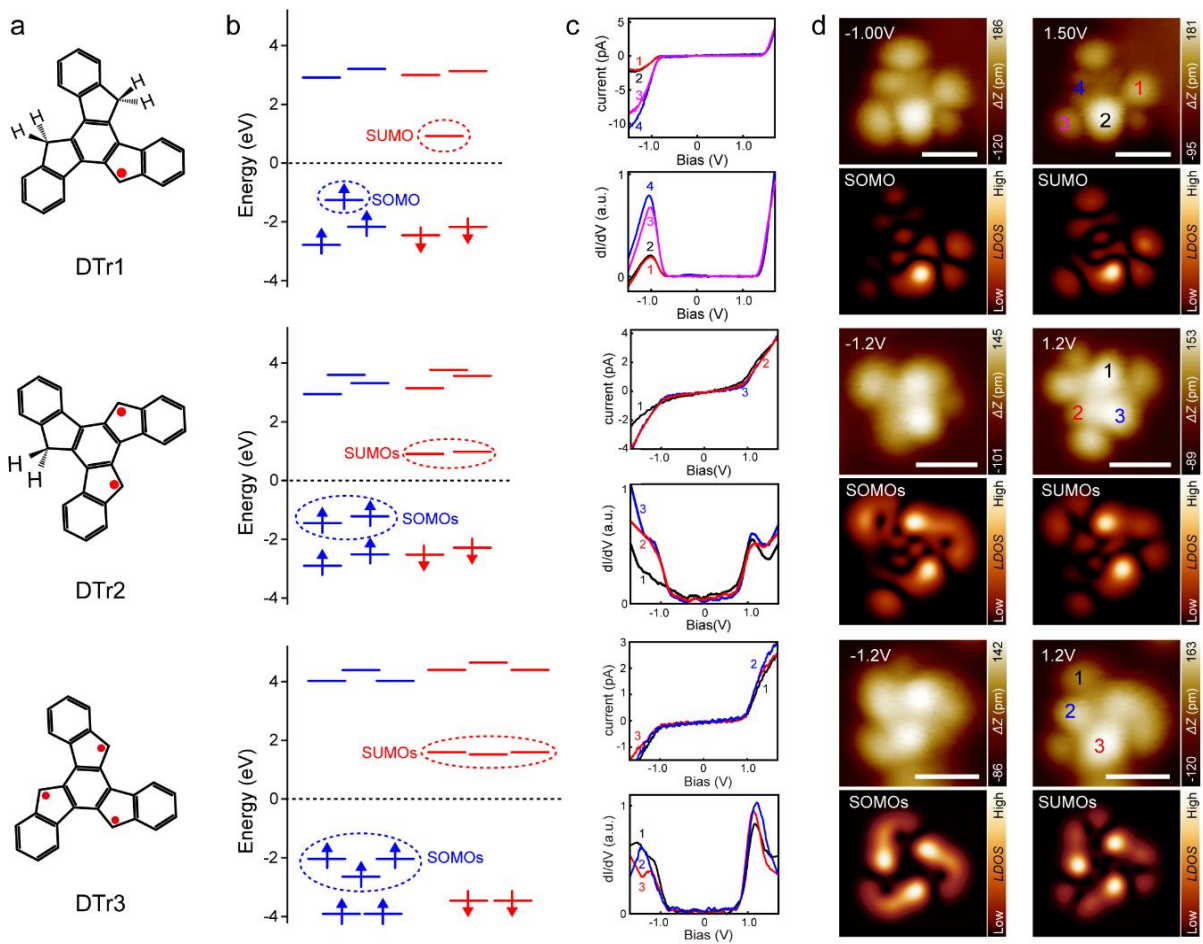

**Figure 4.** Electronic properties of DTrs on a bilayer NaCl island. (a) Chemical structures of DTr1, DTr2, and DTr3. (b) The corresponding spin-polarized DFT calculated energy spectrum of DTr1, DTr2, and DTr3. (c) I/V spectra and corresponding numerical dI/dV spectra of DTr1, DTr2, and DTr3, respectively. The positions are marked in (d). (d) Constant current images (Current: 3 pA; Scale bars: 1 nm.) and the corresponding simulated LDOS images of DTr1, DTr2, and DTr3.

## Conclusion

We demonstrate the ability to realize collective high spin states in molecular graphenoids with regular organized topological defects. Combined with DFT calculations and scanning probe measurements, the presence of collective high spin states of DTrs on a thin insulating film has been confirmed by resolving the SOMOs/SUMOs. Our research not only paves the way for engineering collective high spin states induced by topological defects but provides a new platform for efficient fabrication of ferromagnetic frameworks, such as one-dimensional spin chains and two-dimensional networks with implications for exploring quantum magnetism as well as for quantum technological implications.

## Acknowledgements

S. W. acknowledges the financial support from National Key R&D Program of China (No. 2020YFA0309000), the National Natural Science Foundation of China (No. 11874258, No. 12074247), the Shanghai Municipal Science and Technology Qi Ming Xing Project (No. 20QA1405100), Fok Ying Tung Foundation for young researchers and SJTU (No. 21X010200846). This work is also supported by the Ministry of Science and Technology of China (Grants No. 2019YFA0308600, 2016YFA0301003, No. 2016YFA0300403), NSFC (Grants No. 21925201, No. 11521404, No. 11634009, No. 92065201, No. 11874256, No. 11790313, and No. 11861161003), the Strategic Priority Research Program of Chinese Academy of Sciences (Grant No. XDB28000000) and the Science and Technology Commission of Shanghai Municipality (Grants No. 20ZR1414200, No. 2019SHZDZX01, No. 19JC1412701, No. 20QA1405100) for partial support.

**Keywords:** truxene • nanographenes • high-spin • scanning probe microscopy • surface chemistry

*Supporting Information for*

# Topological Defects Induced Collective High-Spin Quartet State in Truxene-Based Molecular Graphenoids

Can Li,[†][a] Yu Liu,[†][a] Yufeng Liu,[†][a] Fu-Hua Xue,[2][b] Dandan Guan,[a,c,d] Yaoyi Li,[a,c,d] Hao Zheng,[a,c,d] Canhua Liu,[a,c,d] Jinfeng Jia,[a,c,d] Pei-Nian Liu,[b] Deng-Yuan Li,*[b] Shiyong Wang*[a,c,d]

[a] C. Li, Y. Liu, Y. Liu, D. Guan, Y. Li,   Prof. H. Zheng, Prof. C. Liu, Prof. J. Jia, Prof. S. Wang
Key Laboratory of Artificial Structures and Quantum Control (Ministry of Education), Shenyang National Laboratory for Materials Science, School of Physics and Astronomy, Shanghai Jiao Tong University, Shanghai 200240, China
E-mail: shiyong.wang@sjtu.edu.cn

[b] F.-H. Xue, Prof. P.-N. Liu, Dr. D.-Y. Li
Key Laboratory for Advanced Materials and Feringa Nobel Prize Scientist Joint Research Center, Frontiers Science Center for Materiobiology and Dynamic Chemistry, School of Chemistry and Molecular Engineering, East China University of Science & Technology, 130 Meilong Road, Shanghai, 200237, China
E-mail: dengyuanli@ecust.edu.cn

[c] D. Guan, Y. Li,   Prof. H. Zheng, Prof. C. Liu, Prof. J. Jia, Prof. S. Wang
Tsung-Dao Lee Institute, Shanghai Jiao Tong University, Shanghai, 200240, China

[d] D. Guan, Y. Li,   Prof. H. Zheng, Prof. C. Liu, Prof. J. Jia, Prof. S. Wang
Shanghai Research Center for Quantum Sciences, Shanghai 201315, China

[†] These authors contributed equally to this work.

# Table of Contents

# 1. Methods

**STM measurement**

Sample preparation and characterization were carried out using a commercial low-temperature Unisoku Joule-Thomson scanning probe microscopy under ultra-high vacuum conditions (3 x $10^{-10}$ mbar). Au(111) single-crystal was cleaned by cycles of argon ion sputtering, and subsequently annealed to 800 K to get atomically flat terraces. The NaCl were thermally deposited at 800K on clean Au(111) at room temperature for 2mins. Afterwards, the sample was transferred to a cryogenic scanner at 4.9 K for cooling and then truxene were thermally deposited on the clean Au(111) surface with NaCl islands at 10 K. Carbon monoxide molecules are dosed onto the cold sample around 7 K (4 x $10^{-9}$ mbar, 2 minutes). To achieve ultra-high spatial resolution, CO molecule is picked up from Au surface to the apex of tungsten tip. A quartz tuning fork with resonant frequency of 26 KHz has been used in nc-AFM measurements. A lock-in amplifier (589 Hz, 0.1-0.5 mV modulation) has been used to acquire dI/dV spectra. The spectra were taken at 4.9 K unless otherwise stated.

**Spin-polarized DFT calculations**

The ground state structures of gas-phase DTrs were optimized by the PBE0-D3 (BJ)[1-3] functional combined with the def2-SVP basis set,[4] which was extended to a def2-TZVP basis set[4] for the single point energy calculation. Molecular orbitals and electron spin densities were analyzed by Multiwfn[5] and VMD.[6] Images of the structures and isosurfaces were plotted using VESTA.[7]

## 2. Synthesis of truxene precursor

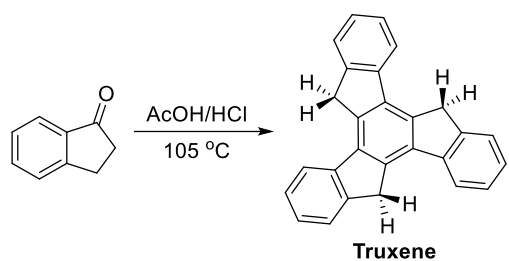

A mixture of 1-indanone (1.9 mmol, 250 mg) in the mixture of acetic (1.4 mL) and concentrated aqueous hydrochloric acids (0.7 mL) was stirred at 105 °C (heating mantle temperature) in a sealed tube under argon atmosphere. After 12 h, the resulting mixture was cooled down to room temperature and then diluted with ice water. The off-white powder was filtered and washed with methanol. The crude product was purified by recrystallization from DCM/*n*-hexane to afford the white solid product truxene (155 mg, 72%). The $^1$H NMR of the obtained product is consistent with literature.[8]

**$^1$H NMR (400 MHz, CDCl$_3$, 25 °C):** δ 7.97 (d, *J* = 7.48 Hz, 3H), 7.71 (d, *J* = 7.44 Hz, 3H), 7.51 (t, *J* = 7.44 Hz, 3H), 7.40 (t, *J* = 7.36 Hz, 3H), 4.30 (s, 6H).

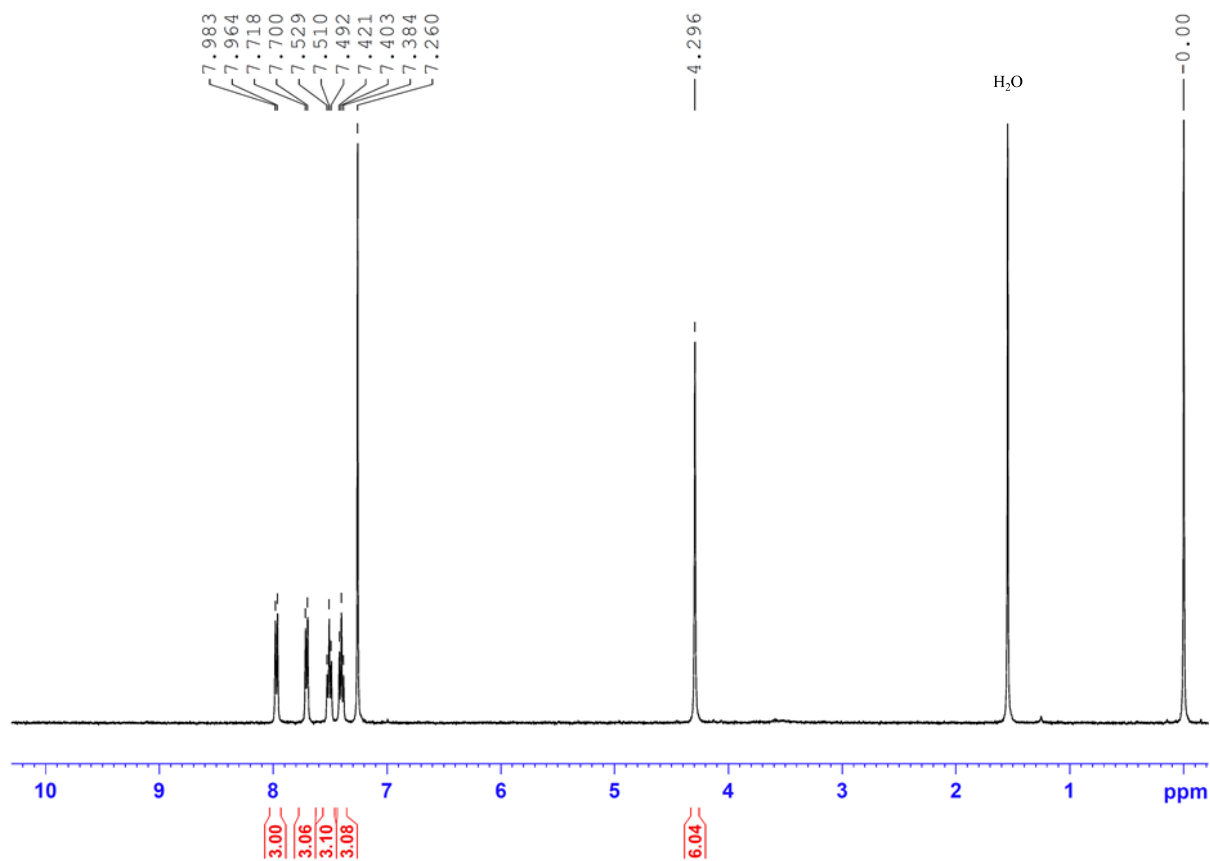

**Figure S1.** $^1$H NMR (400 MHz, CDCl$_3$, 25 °C) spectrum of truxene.

**3. STM tip manipulation for the dehydrogenation of truxene on Au(111)**

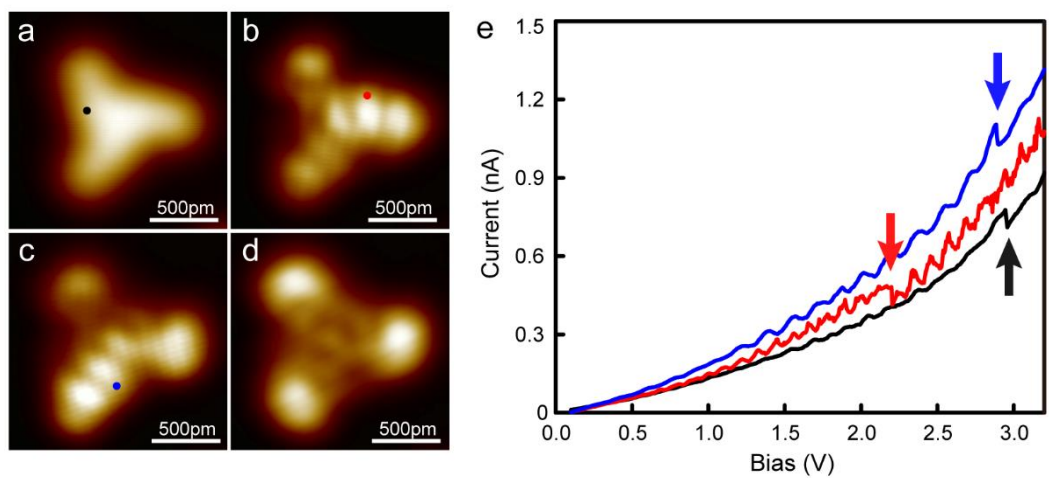

**Figure S2.** STM tip manipulation for the dehydrogenation of truxene on Au(111). (a-d) Constant-height current images of intact truxene molecule, DTr1, DTr2 and DTr3, respectively (bias voltages are 100 mV in (a) and 10 mV in (b-d)). (e) The I-V curves for removing H atoms of DTrs on Au(111) step by step through a slowly ramping bias voltage to 3.2 V with tip raising 400 pm on the molecules corresponding to the positions labelled at (a-d), respectively.

## 4. dI/dV mappings and simulated LDOS images of DTr3 on Au(111)

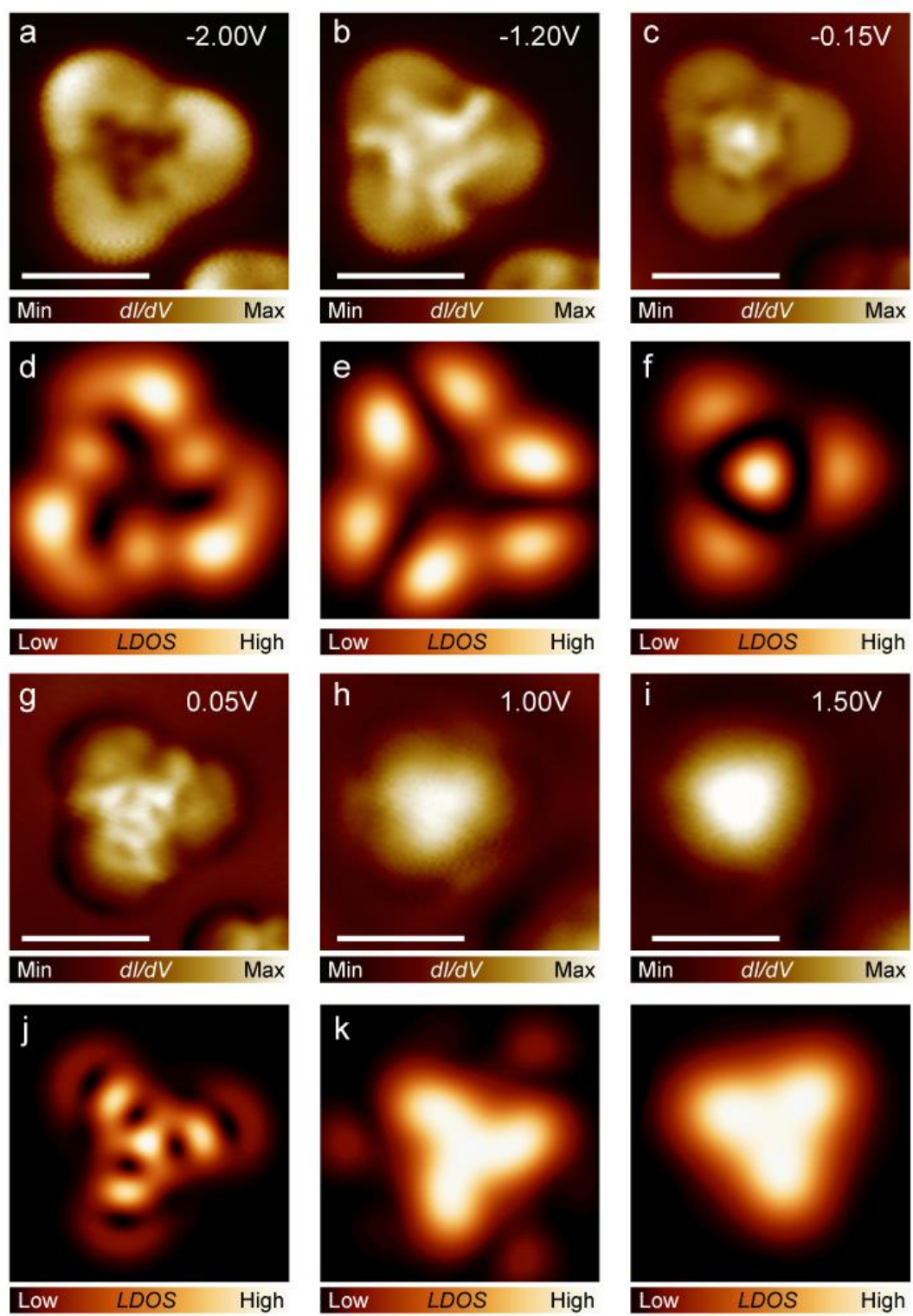

**Figure S3.** dI/dV mappings and simulated LDOS images of DTr3 on Au(111). (a-c and g-i) Experimental dI/dV mappings of DTrs on Au(111). Scale bars: 1 nm. (d-f and j-l) Simulated LDOS images using molecular orbitals of truxene precursor. The dI/dV mappings of dehydrogenated truxene coincide with simulated LDOS images using orbitals from undehydrogenated truxene, indicating the radicals are quenched by Au substrate.

## 5. dI/dV spectra of DTrs on Au(111)

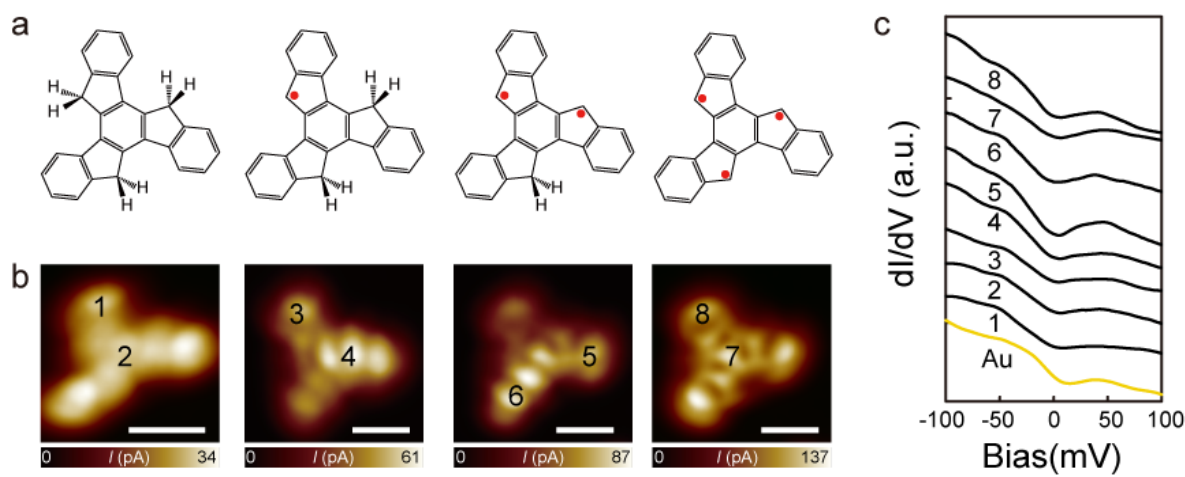

**Figure S4.** Low-energy dI/dV spectra of DTrs on Au(111). (a-b) The chemical structures and constant height current images of DTrs on Au(111). Scale bars: 500 pm. (c) dI/dV spectra taken at locations marked in (b). Kondo resonance has not been observed, indicating the absence of net spins in DTrs.

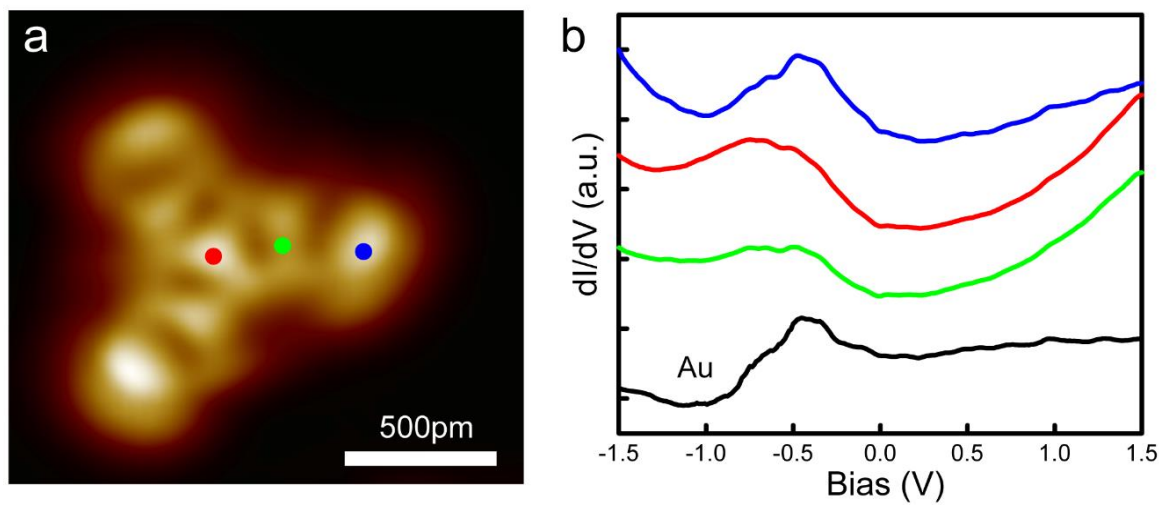

**Figure S5.** Large scale energy dI/dV spectrum of DTr3 on Au(111). (a) Constant-height current images (Bias voltage: 3 mV) of DTr3 on Au(111) with CO-coated tip. (b) Large scale energy dI/dV spectrum of DTr3 on Au(111), which show no obviously feature on this molecules. The blue, red and green line measured at the corresponding position on (a) and the black line is dI/dV spectrum on Au surface.

## 6. STM image showing truxuene molecules on NaCl

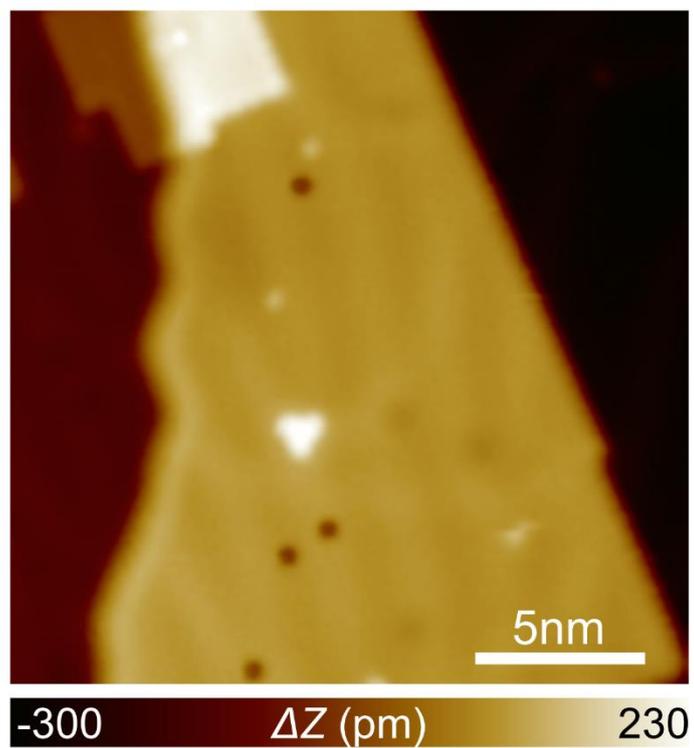

**Figure S6.** STM image (Bias: 1.5 V, Current: 3 pA) showing a truxene precursor on NaCl bilayer island. The black dots are CO molecules.

## 7. Open-shell and closed-shell electronic structure of DTr2

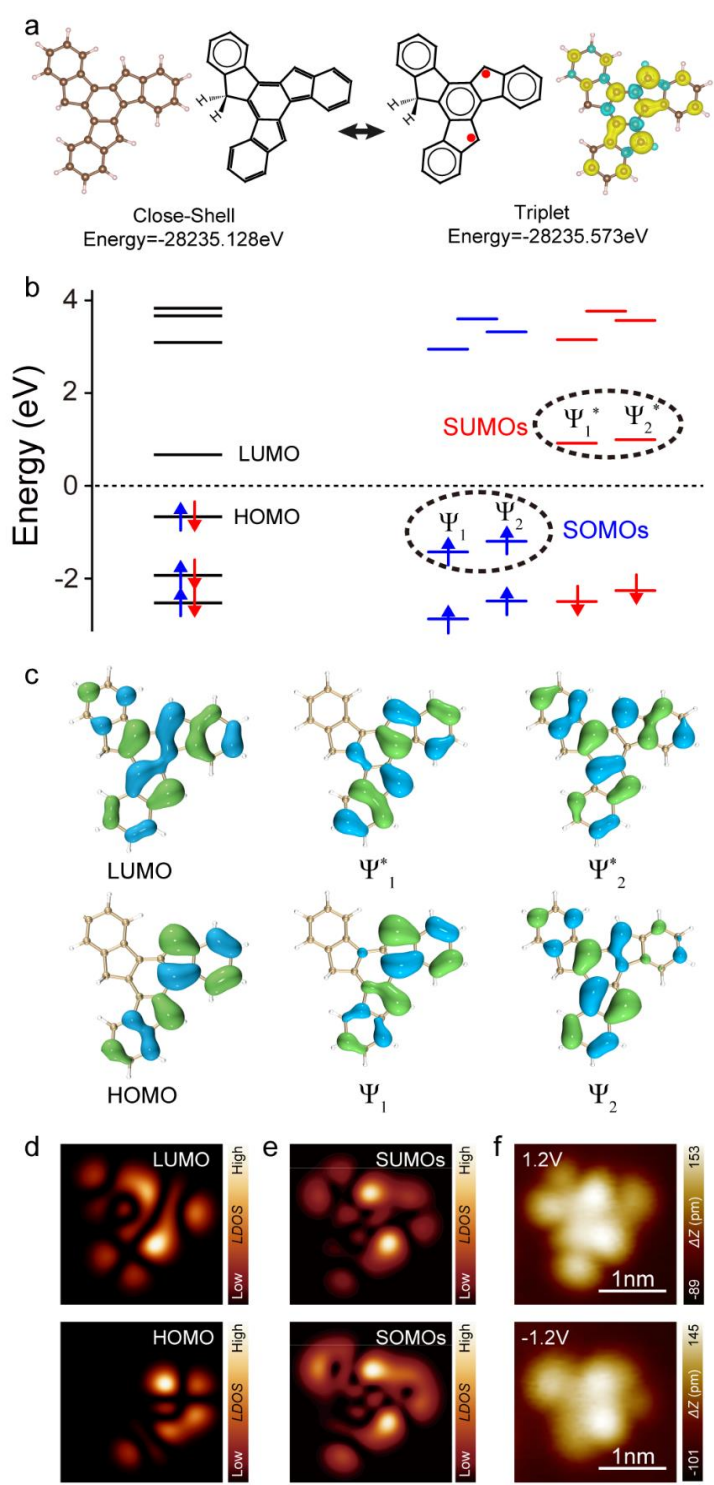

**Figure S7.** Open-shell and closed-shell electronic structure of DTr2. (a) Left: molecular model and chemical structure of closed-shell DTr2. Right: chemical structure and spin density distribution of open-shell DTr2. (b) Energy spectra of DTr2 with closed-shell and open-shell electronic structure. (c) Molecular orbital wavefunctions of DTr2 marked in (b). (d-e) Simulated STM maps of closed-shell and open-shell DTr2. (f) STM images of SUMOs and SOMOs. Experiment STM images coincide with simulated STM images using open-shell frontier orbitals, indicating DTr2 with a triplet ground state.

## 8. Doublet and quartet state of DTr3

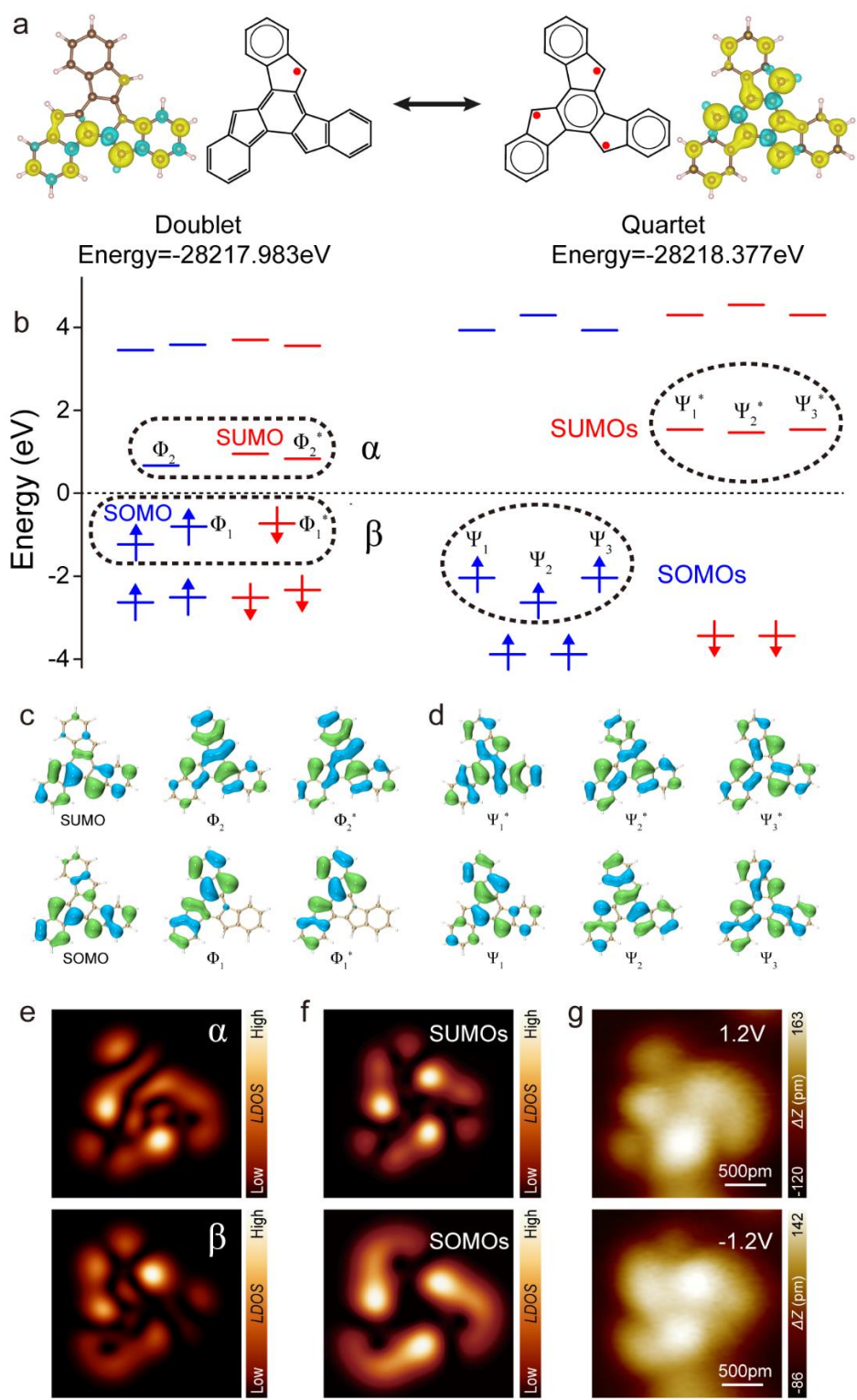

**Figure S8.** Doublet and quartet state of DTr3. (a) Spin density distributions and chemical structures of DTr3 with doublet and quartet state. (b) Energy spectra of DTr3 with doublet and quartet state. (c-d) Molecular orbital wavefunctions of DTr3 marked in (b). (e-g) Simulated STM images of DTr3 with doublet and quartet state. (f) STM images of SUMOs and SOMOs. Experiment STM images coincide with simulated STM images using quartet frontier orbitals, indicating DTr3 with a quartet ground state.

S10